# Solution-processed silver sulphide nanocrystal film for resistive switching memories

Beatriz Martín-García,*† Davide Spirito,*† Roman Krahne and Iwan Moreels‡

Istituto Italiano di Tecnologia, Via Morego 30, 16163 Genova (Italy)
† These authors contributed equally.
‡ Present address: Department of Chemistry, Ghent University. Krijgslaan 281-S3, 9000 Gent, Belgium

Resistive switching memories allow electrical control of the conductivity of a material, by inducing a high resistance (OFF) or a low resistance (ON) state, using electrochemical and ion transport processes. As alternative to high temperature and vacuum-based physical sulphurization methods of silver (Ag), here we propose, as resistive switching medium, a layer built from colloidal $Ag_{2-x}S$ nanocrystals -compatible with solution-processed approaches. The effect of the electrode size (from macro- to micro-scale), composition (Ag, Ti and Pt) and geometry on the device performance together with the electrochemical mechanisms involved are evaluated. We achieved an optimized Ag/Ti bowtie proof-of-concept configuration by e-beam lithography, which fulfils the general requirements for ReRAM devices in terms of low power consumption and reliable $I_{ON}/I_{OFF}$ ratio. This configuration demonstrates reproducible switching between ON and OFF states with data endurance of at least 20 cycles; and an $I_{ON}/I_{OFF}$ ratio up to $10^3$ at low power consumption (0.1V readout), which outperforms previous results in literature for devices with resistive layers fabricated from silver chalcogenide nanoparticles.

## Introduction

Resistive switching random access memories (ReRAMs) are promising nonvolatile memory devices due to their scalability, fast operation response and low power consumption, with applications in information storage technology, or in the development of neuronal networks and logical devices.[1–3] ReRAMs are based on the ability to reversibly switch the resistance of a material between a high resistance state (HRS) and a low resistance state (LRS). General figure of merits for the next generation of ReRAMs are a read voltage between 0.1 and 0.2V, a resistance ratio ($R_{ON}/R_{OFF}$) from 1.2 to 10 for reliable read-out, long endurance (number of write cycles superior to $10^3$, keeping reliable read-out), and data retention longer than 10 years.[3–6]

Usually a ReRAM device consists of a resistive material layer (10 nm – 1μm thickness) between two metal electrodes. A bias voltage above a certain threshold allows to change the resistance state of the material.[1,3] In the particular case of electrochemical metallization (ECM) systems,[1,3] composed by at least one electrochemically active metal, the resistance switching is promoted by redox processes and ion migration, in which the LRS and HRS are related to the formation/growth and dissolution of a conductive "filament" between the metal electrodes.[3,5,7,8] Therefore, compounds with high ion mobility have been proposed for the resistive layers,[7,9] such as silver and copper chalcogenides (e.g. $Ag_2S$, AgGeSe, $Cu_2S$).[4] Among them, $Ag_2S$ stands out as model system due to its low redox overpotential and high mobility at room temperature.[1,3,10–12] The conventional techniques for $Ag_2S$ film fabrication for ReRAMs are physical methods based on Ag sulphurization[12–17] such as chemical vapor deposition, chemical bath deposition and thermal evaporation. These fabrication methods require high vacuum, high temperatures and/or controlled atmosphere. Therefore, the development of more simple alternative methods can open new opportunities in this field, and enable low-cost and scalable device fabrication. Colloidal synthesis methods are a viable option in this respect, offering the possibility to tune and control the atomic composition of the nanocrystals by combining two or more metal precursors, applying partial cation or anion exchange reactions, or forming multishell or dot-in-rod heterostructures.[18–20] Indeed, colloidal nanocrystals (NCs) can be potentially applied as building block for the fabrication of the solid conductive electrolyte, since naturally they present defects and grain boundaries that are known to increase ionic mobility, favoring the switching mechanism.[1,3] Even more, the incorporation of nanoscale "metal inclusions" (*ca.* 5 nm) have been considered as a disruptive technology in memristive devices.[8] Earlier demonstrations using nanoparticles have been reported in the literature, focused on valence-change memory (VCM) devices comprising metal oxides[21–28] or nonvolatile charge-trap memory devices involving quantum dots,[29–38] as composites with polymers or mixed with other molecules. However, less attention has been paid to ECM devices, and only scarce examples can be found, such as sintered $Ag_2Se$ nanocrystal films[39] or $Ag_2Se$-MnO nanoparticles composites[40], despite the broad range of chemical routes that have been proposed for silver chalcogenide NC synthesis[41–44].

Among the Ag chalcogenides (S, Se and Te), $Ag_2S$ stands out due to its higher ion[45] and electron[46] conductivity. Therefore, by choosing colloidal $Ag_2S$ NCs as starting point, we demonstrate a new avenue for accessible resistive materials for ECM-ReRAMs. Synthesis at mild conditions, combined with spin-coating/solid ligand exchange approaches at room temperature, were used for the preparation of the resistive layer in the device consisting of a close-packed $Ag_2S$ NCs film, giving a methodology compatible with solution-processed approaches toward low-cost and large-scale fabrication of ReRAMs. By fabricating lateral, macro-scale systems consisting of two electrodes deposited on the $Ag_2S$ NC-based film, we demonstrate their feasibility and gain insight into the working electrochemical mechanisms involved, while by means of electron beam lithography (EBL) we explore pathways for miniaturization and optimization of electrode design towards proof-of-principle devices. Since for a memristive device an active electrode (*e.g.* Ag), working as metal ions reservoir, together with a second active, or even inert electrode (*e.g.* Pt) are required,[1,3,7] we tested a symmetric configuration with active metal electrodes of Ag, and two asymmetric configurations with Ag as active, and Ti and Pt as inert electrodes. The obtained figures of merit show reproducible and reliable memory characteristics in terms of $I_{ON}/I_{OFF}$ ratio up to $10^3$ at a 0.1V readout and a stable data endurance of at least 20 cycles.

## Results and discussion

In this study, the ion-electron conducting layer is a compact $Ag_{2-x}S$ NCs film prepared by spin coating. The $Ag_{2-x}S$ NCs are prepared by colloidal synthesis from $AgNO_3$ and sulfur powder at 115°C, which can be performed at lower temperature, and does not require high vacuum conditions, as is the case for the established Ag sulphurization physical methods[12–15]. As result, we obtained $Ag_{2-x}S$ NCs with 3 nm diameter and acanthite crystal structure probed by TEM and XRD measurements, with x=0.6 obtained by elemental analysis (see **Electronic Supplementary Information** (ESI) **Figures S1-S2** for NC characterisation). Since the colloidal synthesis yields $Ag_{2-x}S$ NCs that are surrounded by non-conductive organic ligands (oleylamine and oleic acid), an iodine solid ligand exchange is performed during the preparation of the film, leading to a 20 nm thick $Ag_{2-x}S$ NC conductive layer. Briefly, after spin coating the NC film on the substrate, few drops of an iodine salt solution in MeOH are deposited on top and left for 20 s to allow the ligand exchange, followed by a washing step with MeOH. Subsequently, the electrodes consisting of Ag, Ti, or Pt are deposited by e-beam metal evaporation on top of the $Ag_{2-x}S$ NC film.

As starting point for the understanding of and display the working mechanism, we considered an $Ag/Ag_{2-x}S/Ag$ device in lateral configuration.[15] We fabricated simple devices consisting of pairs of square Ag electrodes with size of 1x1 mm$^2$, separated by a gap of about 100µm, by metal deposition *via* a shadow mask on top of the $Ag_{2-x}S$ NC film spin-coated on glass substrates (**Fig. 1a,** see also **ESI Fig. S3a** for an optical microscope image of the complete device). Hereafter, we refer to these devices as 'macroscale systems'. By applying successive DC voltage ramps from 0 to ±2V with a ramp speed of 0.1 Vs$^{-1}$, the formation of metallic Ag protrusions inside the $Ag_{2-x}S$ layer can be observed by scanning electron microscopy (SEM) (**Fig. 1**, see also **ESI Fig. S3a,c** for additional optical microscopy images). In the macroscale systems, the growth process of the metallic Ag protrusions can be followed in real time *in situ* by optical microscopy while performing the DC voltage sweeps, (see **ESI** for movies). We observed the formation of metallic Ag protrusions (**Fig. 1b-e**) that grow from the negatively biased electrode (commonly referred as cathode for ECM devices[47]), and which are partially dissolved when the voltage bias is reversed. At the same time, Ag depletion of the positively biased electrode (noted as anode) occurs (**Fig. 1a,c** and **ESI Fig. S3b**). The Ag protrusions grow preferentially in regions of high electric field, being clearer under prolonged electric field application, when we observed the formation of dendrites that are preferentially oriented along the field lines (see **ESI**, **Fig. S3c**). Elemental analysis by SEM/energy dispersive X-Ray spectroscopy (EDX) indicated that the protrusions and dendrites consist of metallic Ag (**ESI Fig. S4**). The sulphur signal[48] was detected solely from the underlying $Ag_{2-x}S$ film. The formation of metallic Ag under exposure to an electric field can be explained by the occurrence of the redox reaction [$Ag^+$ ($Ag_{2-x}S$) + $e^-$ ↔ $Ag^0$] and migration of the $Ag^+$ ions in the $Ag_{2-x}S$ film towards the cathode, as expected for this system.[3,7,15] This is in accordance with previous results in an Ag/Ag$_2$S/W system on the nanoscale, which investigated such processes by atomic resolution techniques, such as HR-TEM, highlighting that this phenomenon is responsible for the resistive switching.[14]

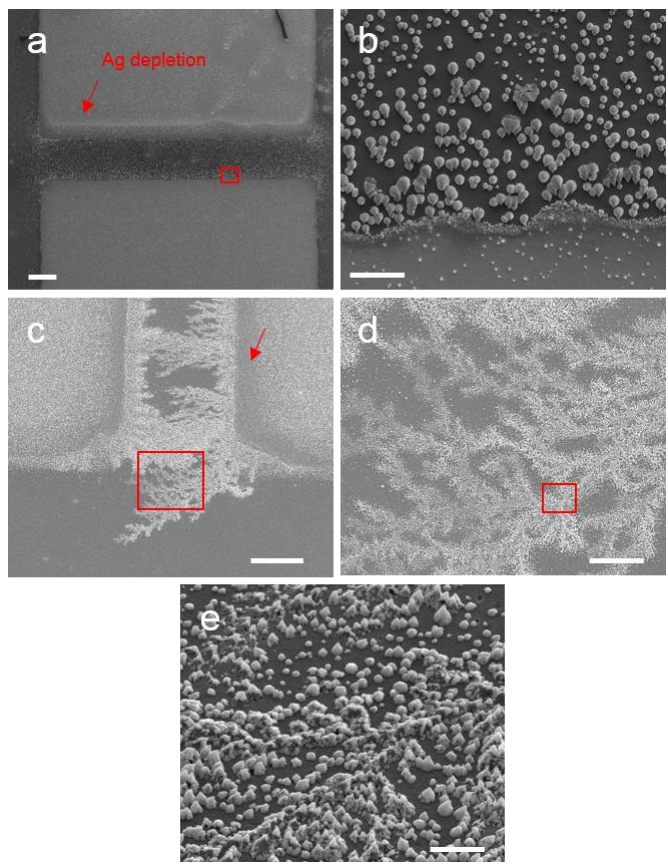

Fig. 1. SEM images of representative Ag/Ag2-xS/Ag macroscale devices at different magnification that have been exposed for short time (a,b) and more extended time (c-e) to an electric field. Scale bars are 100 µm (a), 10 µm (b), 50 µm (c), 10 µm (d) and 1 µm (e). Red square in panel a (c, d) indicates the area shown in panel b (d, e). The depletion of Ag from the top electrode in (a) and formation of Ag protrusions at the bottom electrode (b) are observed, as well as the formation of dendrites along the electric field lines in (c). The SEM image in panel 'e' has been recorded under an angle of 40° with respect to the normal of the substrate surface.

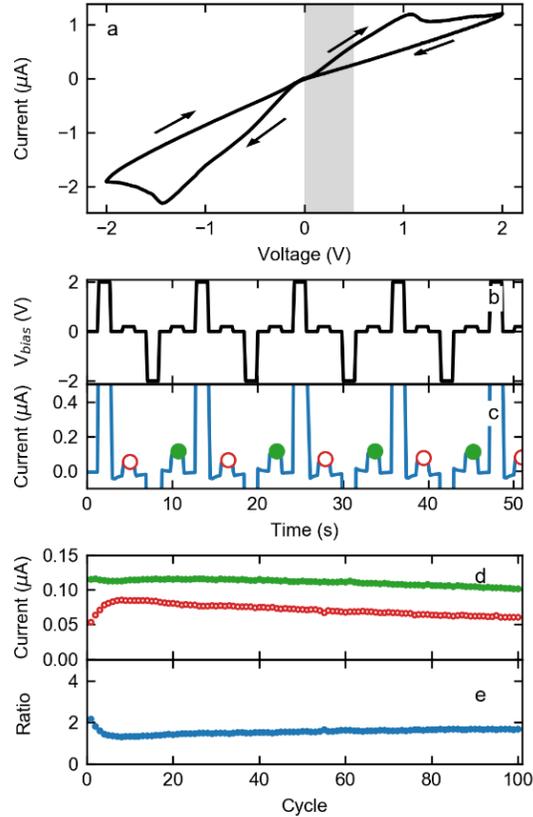

**Fig. 2a** shows the current-voltage (I-V) characteristics from a representative Ag/Ag$_{2-x}$S/Ag macroscale device in the range from -2 V to 2V. Distinct features at 1.1V in the forward sweep (-2V to 2V) and at -1.5 V in the backward sweep (2V to -2V) can be observed, corresponding to the Ag$_2$S/Ag resistive switching established in previous studies, at a voltage around 1.5 V.[49] Control experiments under vacuum do not show any additional features; similar Ag protrusions are observed (see **ESI**, **Fig. S5a** and **c**). Following the hysteresis curve along the directions indicated by arrows, we can distinguish two loops, one under positive, and one under negative bias. In the forward sweep under positive bias starting at 0V, the device goes into a low resistance state. The current first increases almost linearly and then the system enters a region of negative differential resistance when the bias exceeds 1.1 V, corresponding to the formation of Ag protrusions and their extension from the cathode to the anode, as result of the $Ag^+$ ions transport and reduction ($Ag^+$ + $e^-$ → $Ag^0$). In the following backward sweep, the conductivity is markedly lower compared to the preceding forward sweep, except for a small region around 0V. When the bias polarity changes, the current again increases

**Fig. 2** (a) I-V cycle for a representative Ag/Ag$_{2-x}$S/Ag macroscale device, where the shaded area indicates the operational voltage range for the memristor read-out. Time trace of voltage (b) and current (c) during the set/read/reset/read procedure. The dots mark the current value of each read pulse used for calculating the $I_{ON}/I_{OFF}$ ratio. Set/reset voltage values are -2 V and +2 V, respectively, and read voltage is 0.2 V (read pulse 1s). The corresponding current values for ON (●) and OFF (○) states, acquired over 100 set/reset cycles (d) are shown together with the $I_{ON}/I_{OFF}$ ratio (e).

(in absolute value) almost linearly, and then passes through a region of negative differential resistance between -1.5 and -2 V, ascribed to the growth of silver protrusions, now from the opposite electrode, while inducing the dissolution/breaking of the protrusions previously formed on the other electrode. The negative resistance observed from -1.5 V is the result of this balance of formation/dissolution processes. In the forward sweep, the current is again smaller than in the backward sweep. This behaviour corresponds to the 'complementary switching' already reported in Ag$_2$S/Ag devices[17] and other ReRAMs[50–52], related to a limited reservoir of mobile ionic species. In our case, one of the factors that can limit the $Ag^+$ ion availability is the limited supply of Ag metal from the electrode. Indeed, we observe a partial depletion at the interface between metal electrode and the channel (**Fig. 1a,c**), indicating that the electrode does not participate in its entirety to the reservoir of moving species. The dynamics of this phenomenon can be also observed in the movies in the **ESI**. Moreover, we note that the complementary mechanism has a positive threshold voltage ($V_{t,p}$), and a negative threshold voltage ($V_{t,n}$), that determine the onset of the LRS (corresponding to the formation of the filament) for positive and negative bias, respectively.

The hysteresis behaviour in electrical conductivity reported in Fig. 2a can be exploited for memristor functionality, allowing to select appropriate voltage values for the programmable memory. While the observed complementary switching can be useful in integrated memory applications,[51,53] it generally lacks the possibility of read-out at small voltage. In our devices, we observe a hysteresis at low voltage that we exploit for readout at V<0.5V. Considering $Ag_2S$/Ag resistive switching at 1.1 V and -1.5 V, we established voltage bias pulses of 1s at -2V and 2V for the set and reset operations, consisting of Ag protrusions formation (LRS) and their dissolution/break (HRS), respectively. Read-out voltage was chosen in the range from 0.1 V to 0.4 V (marked with a shaded area in Fig. 2a). For much higher set/reset voltages up to ± 10 V, as have been proposed to induce the growth of conductive filaments,[16] we found an increased instability in the switching behaviour in our devices (see **ESI**, control experiments at ±5V, **Fig. S6**).[54] Concerning read-out, the low voltages of 0.1V to 0.4V are already in a range of distinct conductivity switching (at 0.5 V the ratio between HRS and LRS is about 2) and assure compatibility with low-power consumption requirements. **Fig. 2b** shows a representative Ag/$Ag_{2-x}$S/Ag memristor operation over time. Here one cycle consists of a read pulse of 0.2V, followed by a set pulse of -2V, another read pulse of 0.2V, and a reset pulse of 2V. **Fig. 2c** demonstrates that the current trace follows the applied voltage pattern well, and the red and green dots illustrate the read out of the HRS and LRS, respectively. **Figures 2d-e** show the read out current after set (green) and reset (red) pulses, and their ratio (Fig. 2e) for up to 100 cycles.

In the initial cycles, the conductivity of the HRS increase ($I_{OFF}$), while the LRS ($I_{ON}$) is constant. This leads to a decrease in the $I_{ON}/I_{OFF}$ ratio, from about 2 to a stable value of 1.6±0.1 after 8 cycles (a lower value of 1±0.1 was observed for $V_{read}$ = 0.1 and 0.4V). To shorten the number of cycles to reach a stable performance, a pre-activation process of the device can be applied, usually called 'initial forming operation', which should promote an initialization of the $Ag^+$ ion migration across the $Ag_{2-x}$S film[3,55]. Therefore, we applied a linear voltage ramp from 0 to 2V with 0.1 V/s (**ESI**, protocol of pre-activation **Fig. S7a**). In this case, the stabilization occurs earlier (4$^{th}$ cycle) keeping the same ratio at a read-out of 0.2V (**ESI**, **Fig. S7b**).

To improve the device performance, we explored asymmetric device configurations consisting of one active (Ag) and one inert (Ti or Pt) electrode: Ti/$Ag_{2-x}$S/Ag and Pt/$Ag_{2-x}$S/Ag.

Here the Ti electrode was coated with a thin Au layer (10nm) to prevent its oxidation. In this case, the metal deposition, *via* e-beam evaporation through a shadow mask, was carried out in two consecutive steps. While Pt is commonly used as inert electrode, Ti can be a much more cost efficient alternative material, if it does not take part in the switching mechanism and keeps the performance. When using one inert electrode, initially the formation of metallic Ag occurs at the anode. This is again reflected by the formation of the macroscopic dendrites growing from the Ti or Pt electrode as evidenced in **Fig. 3a**, and supported by elemental analysis (**ESI**, SEM/EDX results in **Fig. S8**) that reveals the formation of metallic Ag protrusions at the Ti(Pt) side ($Ag^+ + e^- \rightarrow Ag$) and the Ag depletion at the Ag electrode (Ag $\rightarrow Ag^+ + e^-$). No detectable dissolution or diffusion of Ti is observed, since the preferential redox involves $Ag^+/Ag^0$, and therefore Ti acts as inert electrode in this Ag/Ag$_{2-x}$S system. **Fig. 3b** shows the current-voltage measurements of Ti/Ag$_{2-x}$S/Ag and Pt/Ag$_{2-x}$S/Ag devices, where we observe a qualitatively similar behaviour as in **Fig. 2a**, which underlines that the same switching processes lie at the origin of the changes in conductivity (see **ESI** for control experiments under vacuum in Ti/Ag and Pt/Ag systems, **Fig. S5b** and **d-e**). As for the Ag/Ag$_{2-x}$S/Ag case, we selected the set/reset and read-out voltages from the analysis of the I-V characteristics. In this case the corresponding I-V sweep cycles (**Fig. 3b**) shows a HRS/LRS ratio of *ca.* 4 for Ti and *ca.* 5 for Pt, respectively, at voltages lower than 0.5 V. Therefore, also in this case, a read voltage between 0V and 0.5V could give a reliable and low power consumption read-out. The inset of Fig. 3b shows the voltage trace during a set and reset cycle with three readout pulses at voltages of 0.1V, 0.2V and 0.4V, to which

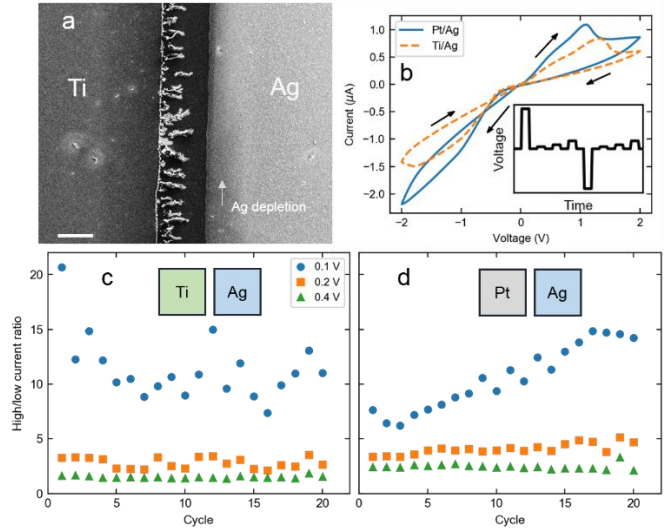

**Fig. 3** (a) Representative SEM image of a Ti/Ag$_{2-x}$S/Ag macroscale device. The Ag depletion region is marked by an arrow. Scale bar is 100 µm. (b) I-V cycles for representative Pt/Ag$_{2-x}$S/Ag (blue solid) and Ti/Ag$_{2-x}$S/Ag (orange dashed) macroscale samples, using the Ag electrode as cathode. The inset shows the 'multi-read' set/reset cycle protocol: reset/set voltage are at +2 V and -2V, respectively, while read voltages are 0.1 V, 0.2 V and 0.4 V. (c, d) I$_{ON}$/I$_{OFF}$ ratio for Ti/Ag$_{2-x}$S/Ag and Pt/Ag$_{2-x}$S/Ag macroscale representative devices, respectively, obtained *via* the 'multi-read protocol': V$_{read}$= 0.1 V (●), 0.2 V (■) and 0.4 V (▲).

we refer hereafter as the 'multi-read protocol'. Similarly as in the Ag/Ag$_{2-x}$S/Ag case, for the Ti/Ag$_{2-x}$S/Ag and Pt/Ag$_{2-x}$S/Ag systems an initial forming operation by a linear ramp from 0 to 2V leads to an earlier stabilization of the I$_{ON}$/I$_{OFF}$ ratio (**Fig. 3c-d** and **ESI** for results without pre-activation (I$_{ON}$/I$_{OFF}$ ratio about 3), **Fig. S7c**). The initialized Ti/Ag$_{2-x}$S/Ag and Pt/Ag$_{2-x}$S/Ag configurations show a small I$_{OFF}$ at 0.1 V that leads to large I$_{ON}$/I$_{OFF}$ ratios of average values of 12±3 and 11±3, respectively, with a reproducible switching endurance of at least 20 cycles. Therefore, the initialization process not only reduces the number of cycles to stabilize the performance, but data suggests that it also can enhance the final I$_{ON}$/I$_{OFF}$ ratio.

The data reported above demonstrate the feasibility of resistive switching in our system with a simple, macroscopic device. However, the figures of merit of the ECM devices can be further improved by optimizing the size and geometry of the device. To this aim, we fabricated micron-scale metal contacts (100x100 µm$^2$, gap 20 µm) *via* e-beam lithography and lift-off techniques on the top of the Ag$_{2-x}$S NC film that was deposited on silicon/SiO$_2$ substrates. To study the effect of geometry, we designed two different electrode shapes, one with a flat front, and one with a triangular tip, and their combinations result in three possible electrode configurations: (i) flat-flat, (ii) flat-tip, and (iii) tip-tip (bowtie) geometry, which are depicted in **Fig. 4a-c**. These geometries were chosen based on reported field simulations[3,15,56] and structural designs[5,13,57], which aim to manipulate the electric field lines and create high field regions. In particular, the tip shape yields regions with high electric field, which drives the dynamics of the filament growth/dissolution, as SEM images demonstrate in **Fig. 4b-c**.

The performance of these devices was evaluated with set/reset cycles using the 'multi-read' procedure, **Fig. 4f-h**, and as before, an initialization process with a linear voltage ramp enhanced the device performance (see **ESI**, **Fig. S7d,** for measurements without initialization). The results obtained from the different geometries and electrode materials are summarised in **Table 1**. From the data, it is clear that the best results are achieved in devices with the tip-tip (bow-tie) design. We observed that even for the tip-tip geometry the Ag/Ag$_{2-x}$S/Ag configuration does not yield an improved $I_{ON}/I_{OFF}$ ratio in the microscale device ($I_{ON}/I_{OFF}$ ratio of 2.2±0.1, **Fig. 4f**), compared to the macroscale device discussed earlier ($I_{ON}/I_{OFF}$ ratio of 1.6). In contrast, devices with Ti and Pt electrodes are significantly improved with an $I_{ON}/I_{OFF}$ ratio of 1011±175 for Ti and 736±82 for Pt, as can be seen in **Fig. 4g-h** (see **ESI**, **Fig. S9a-b**, for other geometries). In this way, comparing the results of 'multi-read' experiments for the different Ag/Ag$_{2-x}$S/Ag, Ti/Ag$_{2-x}$S/Ag and Pt/Ag$_{2-x}$S/Ag microscale devices (**Fig. 4f-h**, see **ESI** for additional 100 cycles tests, **Fig. S10**), clearly the Ti/Ag tip-tip geometry stands out with an $I_{ON}/I_{OFF}$ ratio of about $10^3$ that is achieved at the lowest

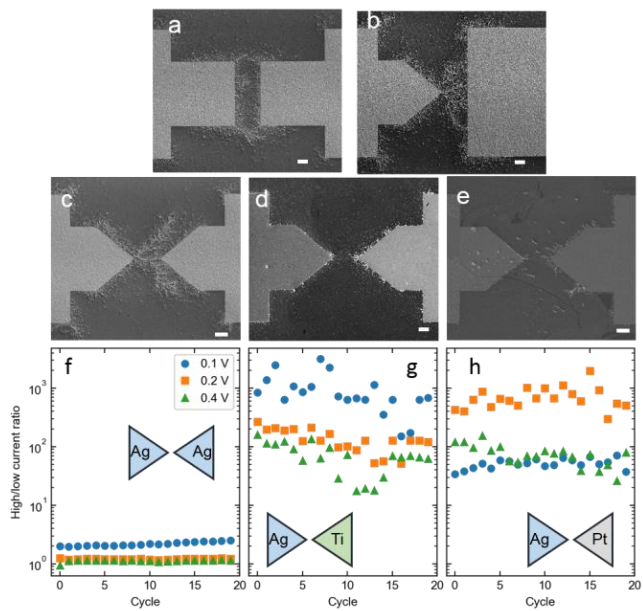

**Fig. 4** (a-e) Representative SEM images of the microscale device geometries fabricated by EBL for the following electrode configurations: flat-flat Ag/Ag (a); tip-flat Ag/Ag (b); and tip-tip Ag/Ag (c), Ag/Ti (d) and Ag/Pt (e). The SEM images were acquired after electrical characterisation. The scale bars are 10µm. (f-h) The corresponding ION/IOFF ratio for tip-tip microscale representative devices using as electrodes Ag/Ag, Ag/Ti and Ag/Pt, shown on top, respectively, with Vread= 0.1 V (●), 0.2 V (■) and 0.4 V (▲). The Ag contact was the cathode (-) in the Ti/Ag2-xS/Ag and Pt/Ag2-xS/Ag systems.

readout voltage of 0.1V. These results demonstrate the feasibility of the straightforward NCs-based fabrication approach here proposed, the relevance of the device design and the potential of the less explored and much more cost efficient Ti material as inert electrode in memristors, that can reduce the material cost[58] of a device by a factor of five with respect to Pt.

**Table 1**. Best $I_{ON}/I_{OFF}$ ratio performance for the different electrode geometries and materials in the macro- and micro-scale devices fabricated by shadow mask and EBL techniques. The readout voltage is reported in parenthesis.

| Electrodes | Geometry | | | | |
|---|---|---|---|---|---|
| | Macroscale | EBL designs | | | |
| Material | flat/flat | flat/flat | flat (Ag) /tip | tip(Ag)/flat | tip/tip |
| Ag/Ag | 1.6±0.1 (0.2 V) | 1.7±0.1 (0.1 V) | 1.9±0.1 (0.1 V) | | 2.2±0.1 (0.1 V) |
| Ti/Ag | 12±3 (0.1 V) | 31±4 (0.1 V) | 651±139 (0.1 V) | 298±60 (0.1 V) | 1011±175 (0.1 V) |
| Pt/Ag | 11±3 (0.1 V) | 115±28 (0.2-0.4 V) | 89±19 (0.2-0.4 V) | 46±4 (0.1 V) | 736±82 (0.2-0.4 V) |

## Conclusions

In conclusion, we explored Ag$_{2-x}$S colloidal NC films for resistive switching devices, studied the electric field induced transformations, and characterised their performance in different device configurations. At the macroscale, we demonstrated the functionality of resistive switching and highlighted the mechanisms involved. Therefore, in this way the best performance is an $I_{ON}/I_{OFF}$ ratio of 15 obtained with the asymmetric Ti/Ag$_{2-x}$S/Ag design, which was already one order of magnitude larger than that of the symmetric Ag/Ag$_{2-x}$S/Ag device. The devices displayed reproducible resistive switching between the ON and OFF states and read-out for at least 100 cycles. This is improved in microscale devices with a gap of around 20 µm, keeping the Ti/Ag$_{2-x}$S/Ag configuration, where the electrodes in a tip-tip (bowtie) geometry enabled an $I_{ON}/I_{OFF}$ ratio enhancement of up to 2 orders of magnitude, with an $I_{ON}/I_{OFF}$ ratio of about $10^3$ during at least 20 cycles at 0.1V read voltage. These results satisfy the basic requirements for reliable devices and show improved performance in comparison with other Ag chalcogenide NC memristors[39]. Our results greatly enhance the HRS/LRS ratio by almost 2 orders of magnitude compared to Ag/Ag$_2$Se/Au NC-based devices (with a ratio of about 30).[39] Compared to Ag$_2$S memristors prepared by physical methods and with an advanced design, our devices are still in an early stage concerning switching speed, but competitive in ON/OFF ratio, where, for example, values of 10 for Pt-Ir/Ag$_2$S/Ag devices,[5] and close to 100 for Pt/Ag$_2$S/Ag devices[12] have been reported. Our results thus highlight that colloidal NCs are a promising material for solution-based fabrication of resistive switching memory devices, keeping in mind that remains room for improvement in terms of device speed and endurance. Future possibilities can be envisioned from the EBL design of the electrodes as already pointed out in this work, such as alternative device architectures (triode switching[48], gap-type 'tug of war' switch[16], three-terminal device[59]), and/or from the colloidal synthesis advantages, concerning the control of material composition, doping or NC shape, which can be explored in future work.

## Experimental

**Materials.** AgNO$_3$ (99.8%), oleic acid (OA, 90%, degassed at 100 °C for 2h), 1-octadecene (ODE, 90%, degassed at 150 °C for 3h), sulfur powder (99.98%), toluene (anhydrous, ≥99.8%), tetrachloroethylene (TCE, anhydrous, ≥99%), acetone (≥99.5%), ethanol (99.8%), methanol (anhydrous, 99.8%), 2-propanol (≥99.8%), tetramethylammonium iodine (TMAI, 99%) were purchased from Sigma-Aldrich and oleylamine (OlAm, C$_{18}$-content 80-90%, degassed) from Acros Organics. The water used in the ICP-OES sample preparation was obtained from a MilliQ® system and the HCl (≥ 37%, TraceSELECT® Fluka) and HNO$_3$ (≥69%, TraceSELECT® Fluka) were purchased from Sigma Aldrich. Glass slides and p-doped silicon wafers (boron doping, 90 nm thermal SiO$_2$, 1-10 Ωcm, University wafer®). Resist (2:1) polymethylmethacrylate (PMMA A4) and anisole (A thinner) supplied by MicroChem®. Ag (99.999%), Ti (99.995%), Pt (99.99%), Cu (99.999%) and Au (99.999%) pellets were purchased from Kurt J. Lesker®.

**Ag$_{2-x}$S nanocrystal synthesis**. The synthesis of colloidal Ag$_{2-x}$S NCs with photoluminescence emission around 1200 nm (see ESI, Fig. S1) was performed based on established protocols, but without using dodecanethiol (DDT)[60,61]. This protocol promotes metallic Ag formation which may enhance electron mobility in the nanocrystal film.[11] We mixed AgNO$_3$ (68 mg), OlAm (1 mL), OA (5 mL) and ODE (10 mL), and heated and degassed under vacuum in a Schlenk line at 95°C for 2h, obtaining a clear yellowish solution. The flask was then heated to 115°C, and a solution of S-OlAm (1 mL, 0.3M) mixed with pure OlAm (3 mL) was rapidly injected. After growing the NCs for 1 min, the synthesis was quickly quenched by cooling the flask to room temperature. The S-OlAm stock solution was prepared dissolving 0.16 g of S in 15 mL of OlAm under vacuum for 30 min at 120°C,[62] and stored in a glove box. The NC suspension, as obtained from synthesis, was purified (in air) by adding acetone, followed by centrifugation, followed by a second washing step with ethanol and drying with a N$_2$ flow. The NC powder was finally dispersed in toluene and filtered with a 0.2 µm

PTFE filter membrane (Sartorius®) obtaining a concentration of 30 mg mL$^{-1}$, which was used for the device fabrication.

**Device fabrication**. The substrates (glass or Si/SiO$_2$) were cleaned in an ultrasonic bath (8 min each step), first with acetone, then followed by isopropanol, and finally dried with a N$_2$ flow. The Ag$_{2-x}$S NC layer (ca. 20 nm, measured by a Veeko Dektak® profilometer) was prepared by spin coating and solid ligand exchange procedures in a glove box. To deposit this layer, the NC dispersion (30 mg mL$^{-1}$, 5-15 µL to cover the substrate) was spin coated at 2500 rpm for 30 s. Then, the film was covered by a TMAI solution in MeOH (1 mg mL$^{-1}$)[63] for 20 s and then rinsed with several drops of MeOH before spin coating at 2500 rpm for 30 s. The substrate was rinsed two more times with MeOH, followed by spinning at 2500 rpm for 30 s to dry the film. The films were stored in a N$_2$-filled box overnight before processing the final device, to ensuring complete solvent evaporation. For the macroscale devices, we used a shadow mask to deposit the metals on 1x1 mm$^2$ areas, separated by gaps of *ca.* 100 µm. In the case of asymmetric electrodes, two consecutive evaporations were performed exposing half of the device each time. The metals were evaporated in a Kenosistec® e-beam evaporator (deposition rate of 0.3 Å.s$^{-1}$ and base pressure of about 1.0 10$^{-6}$ mbar) with thickness of 50 nm (Ag, Ti and Pt); for Ag and Ti an additional layer of 10 nm Au was evaporated to prevent oxidation. Microscale devices were fabricated, on Si substrates with a 90 nm SiO$_2$ top layer, by e-beam lithography (EBL), using spin-coated PMMA as resist (baked at 170ºC on a hot plate). The metal deposition for the electrodes was performed as for the macroscale devices, followed by lift-off in acetone and rising in isopropanol.

Reproducibility was tested by fabricating devices with 3 different Ag$_{2-x}$S NC synthesis batches, used for both macro- and microscale devices. For macroscale devices, at least 5 devices were tested in every configuration. For microscale devices (EBL fabrication), chips with 20-30 devices of each kind were fabricated; at least 10 were tested. The results are comparable between the different sets.

**SEM/EDX characterisation**. We performed SEM/EDX measurements on the devices using a Helios Nanolab 600 (FEI Company) combined with an X-Max detector and INCA® system (Oxford Instruments) for the EDX spectra acquisition and analysis. SEM measurements were performed at 10-20 kV and 0.2 nA, while EDX measurement conditions were set at 10kV and 0.8nA. The SEM measurements were carried out with voltage ≤20 kV and current ≤0.8nA, since in previous studies[14,16] – confirmed by our own observation – Ag protrusions can be formed under high density e-beam exposure. Thus, the focus (brightness, contrast and stigmatism) and beam adjustments were done outside of the imaged area and the photo was collected with an exposure time less than 1µs per pixel.

**Electrical measurements**. Electrical characterisation was performed in air, using a probe station equipped with a Keithley 2612 sourcemeter in two-probe configuration. The high and low terminals are connected to the device. In devices with asymmetric electrodes (*e.g.* Ag/Ti) the low terminal is connected to Ag. The measurement procedure was controlled by a PC using Python® functions (via the Pymeasure package[64]); the data were subsequently handled and analysed using the Python packages Numpy,[65] Scipy,[66] Pandas[67] and Matplotlib[68]. In the set/read/reset measurements, for each cycle the I$_{ON}$ (LRS) and I$_{OFF}$ (HRS) values are acquired by taking the average over the read pulse, avoiding possible spikes during the measurements.

# Conflicts of interest
There are no conflicts to declare.

## Acknowledgements

This project has received funding from the European Union's Horizon 2020 research and innovation programme under grant agreement No 696656 (GrapheneCore1). The authors thank Dr. F. De Angelis for the access to the IIT clean room facilities and M. Leoncini for support during the SEM/EDX measurements and evaporation procedures. We thank Dr. M. Prato for access to the XRD and ICP-OES at the Materials Characterization facilities, and F. Drago for performing the ICP-OES analysis.

## Notes and references

1  R. Waser and M. Aono, *Nat. Mater.*, 2007, **6**, 833–840.
2  F. Pan, S. Gao, C. Chen, C. Song and F. Zeng, *Mater. Sci. Eng. R Rep.*, 2014, **83**, 1–59.
3  R. Waser, R. Dittmann, G. Staikov and K. Szot, *Adv. Mater.*, 2009, **21**, 2632–2663.
4  I. Valov, E. Linn, S. Tappertzhofen, S. Schmelzer, J. van den Hurk, F. Lentz and R. Waser, *Nat. Commun.*, 2013, **4**, 1771.
5  A. Geresdi, M. Csontos, A. Gubicza, A. Halbritter and G. Mihály, *Nanoscale*, 2014, **6**, 2613–2617.
6  R. Waser, D. Ielmini, H. Akinaga, H. Shima, H.-S. P. Wong, J. J. Yang and S. Yu, in *Resistive Switching*, eds. D. Ielmini and R. Waser, Wiley-VCH Verlag GmbH & Co. KGaA, Weinheim, Germany, 2016, pp. 1–30.
7  I. Valov, R. Waser, J. R. Jameson and M. N. Kozicki, *Nanotechnology*, 2011, **22**, 254003.
8  Y. Yang, P. Gao, L. Li, X. Pan, S. Tappertzhofen, S. Choi, R. Waser, I. Valov and W. D. Lu, *Nat. Commun.*, , DOI:10.1038/ncomms5232.
9  M. N. Kozicki and H. J. Barnaby, *Semicond. Sci. Technol.*, 2016, **31**, 113001.
10  H. Schmalzried, *Prog. Solid State Chem.*, 1980, **13**, 119–157.
11  I. Riess, *Solid State Ion.*, 2003, **157**, 1–17.
12  K. Terabe, T. Hasegawa, T. Nakayama and M. Aono, *Nature*, 2005, **433**, 47–50.
13  A. Nayak, T. Tamura, T. Tsuruoka, K. Terabe, S. Hosaka, T. Hasegawa and M. Aono, *J. Phys. Chem. Lett.*, 2010, **1**, 604–608.
14  Z. Xu, Y. Bando, W. Wang, X. Bai and D. Golberg, *ACS Nano*, 2010, **4**, 2515–2522.
15  A. Gubicza, D. Z. Manrique, L. Pósa, C. J. Lambert, G. Mihály, M. Csontos and A. Halbritter, *Sci. Rep.*, , DOI:10.1038/srep30775.
16  C. Lutz, T. Hasegawa and T. Chikyow, *Nanoscale*, 2016, **8**, 14031–14036.
17  P. Cheng and Y. H. Hu, *J. Mater. Chem. C*, 2015, **3**, 2768–2772.
18  D. V. Talapin, J.-S. Lee, M. V. Kovalenko and E. V. Shevchenko, *Chem. Rev.*, 2010, **110**, 389–458.
19  M. V. Kovalenko, L. Manna, A. Cabot, Z. Hens, D. V. Talapin, C. R. Kagan, V. I. Klimov, A. L. Rogach, P. Reiss, D. J. Milliron, P. Guyot-Sionnnest, G. Konstantatos, W. J. Parak, T. Hyeon, B. A. Korgel, C. B. Murray and W. Heiss, *ACS Nano*, 2015, **9**, 1012–1057.
20  A. Teitelboim, N. Meir, M. Kazes and D. Oron, *Acc. Chem. Res.*, 2016, **49**, 902–910.
21  D. O. Schmidt, S. Hoffmann-Eifert, H. Zhang, C. La Torre, A. Besmehn, M. Noyong, R. Waser and U. Simon, *Small*, 2015, **11**, 6444–6456.
22  F. Verbakel, S. C. J. Meskers, D. M. de Leeuw and R. A. J. Janssen, *J. Phys. Chem. C*, 2008, **112**, 5254–5257.
23  M. M. Shirolkar, J. Li, X. Dong, M. Li and H. Wang, *Phys. Chem. Chem. Phys.*, 2017, **19**, 26085–26097.
24  J. Wang, S. Choudhary, W. L. Harrigan, A. J. Crosby, K. R. Kittilstved and S. S. Nonnenmann, *ACS Appl. Mater. Interfaces*, 2017, **9**, 10847–10854.
25  J. Wang, S. Choudhary, J. De Roo, K. De Keukeleere, I. Van Driessche, A. J. Crosby and S. S. Nonnenmann, *ACS Appl. Mater. Interfaces*, 2018, **10**, 4824–4830.
26  J. H. Jung, J.-H. Kim, T. W. Kim, M. S. Song, Y.-H. Kim and S. Jin, *Appl. Phys. Lett.*, 2006, **89**, 122110.
27  F. Li, T. W. Kim, W. Dong and Y.-H. Kim, *Appl. Phys. Lett.*, 2008, **92**, 11906.
28  E. Goren, M. Ungureanu, R. Zazpe, M. Rozenberg, L. E. Hueso, P. Stoliar, Y. Tsur and F. Casanova, *Appl. Phys. Lett.*, 2014, **105**, 143506.
29  F. Li, D.-I. Son, J.-H. Ham, B.-J. Kim, J. H. Jung and T. W. Kim, *Appl. Phys. Lett.*, 2007, **91**, 162109.
30  N. G. Portney, A. A. Martinez-Morales and M. Ozkan, *ACS Nano*, 2008, **2**, 191–196.

31 D. H. Kim, C. Wu, D. H. Park, W. K. Kim, H. W. Seo, S. W. Kim and T. W. Kim, *ACS Appl. Mater. Interfaces*, DOI:10.1021/acsami.7b18817.
32 B. C. Das and A. J. Pal, *ACS Nano*, 2008, **2**, 1930–1938.
33 S. Sahu, S. K. Majee and A. J. Pal, *Appl. Phys. Lett.*, 2007, **91**, 143108.
34 D.-I. Son, J.-H. Kim, D.-H. Park, W. K. Choi, F. Li, J. H. Ham and T. W. Kim, *Nanotechnology*, 2008, **19**, 55204.
35 F. Li, D.-I. Son, S.-M. Seo, H.-M. Cha, H.-J. Kim, B.-J. Kim, J. H. Jung and T. W. Kim, *Appl. Phys. Lett.*, 2007, **91**, 122111.
36 D. Y. Yun, J. H. Jung, D. U. Lee, T. W. Kim, E. D. Ryu and S. W. Kim, *Appl. Phys. Lett.*, 2010, **96**, 123302.
37 Y. C. Ju, S. Kim, T.-G. Seong, S. Nahm, H. Chung, K. Hong and W. Kim, *Small*, 2012, **8**, 2849–2855.
38 D. Y. Yun, T. W. Kim and S. W. Kim, *Thin Solid Films*, 2013, **544**, 433–436.
39 J. Jang, F. Pan, K. Braam and V. Subramanian, *Adv. Mater.*, 2012, **24**, 3573–3576.
40 Q. Hu, T. S. Lee, N. J. Lee, T. S. Kang, M. R. Park, T.-S. Yoon, H. H. Lee and C. J. Kang, *J. Nanosci. Nanotechnol.*, 2017, **17**, 7189–7193.
41 A. Sahu, L. Qi, M. S. Kang, D. Deng and D. J. Norris, *J. Am. Chem. Soc.*, 2011, **133**, 6509–6512.
42 M. Yarema, S. Pichler, M. Sytnyk, R. Seyrkammer, R. T. Lechner, G. Fritz-Popovski, D. Jarzab, K. Szendrei, R. Resel, O. Korovyanko, M. A. Loi, O. Paris, G. Hesser and W. Heiss, *ACS Nano*, 2011, **5**, 3758–3765.
43 B. J. Beberwyck, Y. Surendranath and A. P. Alivisatos, *J. Phys. Chem. C*, 2013, **117**, 19759–19770.
44 D. Ruiz, B. del Rosal, M. Acebrón, C. Palencia, C. Sun, J. Cabanillas-González, M. López-Haro, A. B. Hungría, D. Jaque and B. H. Juarez, *Adv. Funct. Mater.*, 2017, **27**, 1604629.
45 S. Miyatani, *J. Phys. Soc. Jpn.*, 1981, **50**, 3415–3418.
46 C. Xiao, J. Xu, K. Li, J. Feng, J. Yang and Y. Xie, *J. Am. Chem. Soc.*, 2012, **134**, 4287–4293.
47 M. N. Kozicki, M. Mitkova and I. Valov, in *Resistive Switching*, eds. D. Ielmini and R. Waser, Wiley-VCH Verlag GmbH & Co. KGaA, Weinheim, Germany, 2016, pp. 483–514.
48 T. Sakamoto, N. Iguchi and M. Aono, *Appl. Phys. Lett.*, 2010, **96**, 252104.
49 S. Tappertzhofen, R. Waser and I. Valov, *ChemElectroChem*, 2014, **1**, 1287–1292.
50 F. Nardi, S. Balatti, S. Larentis, D. C. Gilmer and D. Ielmini, *IEEE Trans. Electron Devices*, 2013, **60**, 70–77.
51 J. van den Hurk, V. Havel, E. Linn, R. Waser and I. Valov, *Sci. Rep.*, , DOI:10.1038/srep02856.
52 Y. Yang, P. Sheridan and W. Lu, *Appl. Phys. Lett.*, 2012, **100**, 203112.
53 E. Linn, R. Rosezin, C. Kügeler and R. Waser, *Nat. Mater.*, 2010, **9**, 403–406.
54 M. Arita, Y. Ohno, Y. Murakami, K. Takamizawa, A. Tsurumaki-Fukuchi and Y. Takahashi, *Nanoscale*, 2016, **8**, 14754–14766.
55 D. S. Jeong, B. J. Choi and C. S. Hwang, in *Resistive Switching*, eds. D. Ielmini and R. Waser, Wiley-VCH Verlag GmbH & Co. KGaA, Weinheim, Germany, 2016, pp. 289–316.
56 X. Guo, C. Schindler, S. Menzel and R. Waser, *Appl. Phys. Lett.*, 2007, **91**, 133513.
57 N. Onofrio, D. Guzman and A. Strachan, *Nanoscale*, 2016, **8**, 14037–14047.
58 Kurt J. Lesker - Evaporation Materials Price, https://www.lesker.com/newweb/menu_depositionmaterials.cfm?section=evapmats&init=skip.
59 M. Ungureanu, R. Llopis, F. Casanova and L. E. Hueso, *Appl. Phys. Lett.*, 2014, **104**, 13503.
60 P. Jiang and Z. Chen, *New J. Chem.*, 2017, **41**, 5707–5712.
61 L. Motte and J. Urban, *J. Phys. Chem. B*, 2005, **109**, 21499–21501.
62 I. Moreels, Y. Justo, B. De Geyter, K. Haustraete, J. C. Martins and Z. Hens, *ACS Nano*, 2011, **5**, 2004–2012.
63 M. Bernechea, N. C. Miller, G. Xercavins, D. So, A. Stavrinadis and G. Konstantatos, *Nat. Photonics*, 2016, **10**, 521–525.
64 C. Jermain, Minhhai, G. Rowlands, D. Spirito, B. Feinstein, Nowacklab-User, G. A. Vaillant, C. Buchner, T. V. Boxtel, Marc-Alexandre Chan, Mhdg, Ederag, Sriharshava, N. R, M. E. Lund, Julian and G. P, Ralph-Group/Pymeasure, DOI: 10.5281/zenodo.1218272, (accessed May 9, 2018).
65 Travis E, Oliphant, *A guide to NumPy*, Trelgol Publishing, USA, 2006.
66 Jones E, Oliphant E, Peterson P, et al., SciPy: Open Source Scientific Tools for Python, http://www.scipy.org/, (accessed May 9, 2018).
67 W. McKinney, in *Proceedings of the 9th Python in Science Conference*, eds. S. van der Walt and J. Millman, 2010, pp. 51–56.
68 M. Droettboom, T. A. Caswell, J. Hunter, E. Firing, J. H. Nielsen, N. Varoquaux, B. Root, E. S. D. Andrade, P. Elson, D. Dale, Jae-Joon Lee, A. Lee, J. K. Seppänen, R. May, D. Stansby, D. McDougall, A. Straw, P. Hobson, T. S. Yu, E. Ma,

C. Gohlke, S. Silvester, C. Moad, A. F. Vincent, J. Schulz, P. Würtz, F. Ariza, Cimarron, T. Hisch and N. Kniazev, Matplotlib/Matplotlib, DOI: 10.5281/zenodo.1004650.

# Electronic Supplementary Information

# Solution-processed silver sulphide nanocrystal film for resistive switching memories

*Beatriz Martín-García,\*† Davide Spirito,\*† Roman Krahne and Iwan Moreels*

Istituto Italiano di Tecnologia. Via Morego 30, 16163 Genova (Italy)

**$Ag_{2-x}S$ NC characterisation.** *Transmission electron microscopy* (TEM) images were acquired with a 100 kV JEOL JEM-1011 microscope equipped with a thermionic gun. Samples were drop casted onto carbon-coated 200 mesh copper grids for the measurements. From the TEM measurements, an average NC diameter of *ca.* 3 nm was obtained (**Fig. S1a**).

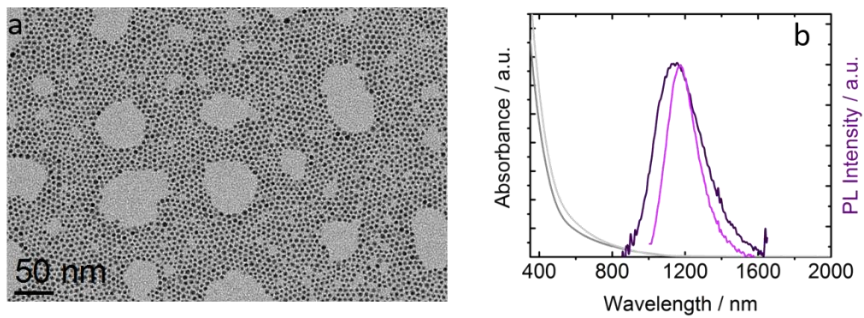

**Fig. S1.** (a) Representative TEM image of $Ag_{2-x}S$ NCs. (b) Absorbance and photoluminescence spectra of two $Ag_{2-x}S$ NC batches dispersed in TCE.

*Absorption spectra* were recorded with a Varian Cary 5000 UV-vis-NIR spectrophotometer. The *steady-state photoluminescence* (PL) emission was collected using an Edinburgh Instruments FLS920 spectrofluorometer, exciting the samples at 400 nm. The absorption spectrum does not show any distinct features. However, the synthesised NCs show a NIR PL emission around 1200 nm, which is in agreement with other $Ag_{2-x}S$ NCs reported in literature.[1,2] As shown in **Fig. S1b**, the PL emission properties are reproducible from batch to batch.

*Elemental analysis* was performed on digested NC solutions by inductively coupled plasma-optical emission spectrometry (ICP-OES). Samples were prepared in a 25 mL volumetric flask, drying a known amount of toluene solution under $N_2$ flow, and digesting the dry residue overnight in two different acid media: (i) in 2.5 mL of aqua regia (HCl:HNO$_3$, 3:1vol) for standard analysis, and (ii) in HNO$_3$ for an improved Ag detection. Prior to the measurements, the samples were diluted to a total volume of 25 mL with Millipore water, and stirred with a vortex mixer for 10 s at 2400 rpm. Then, the sample was filtered using a PTFE membrane (0.45µm pore size, Sartorius®). Measurements were carried out with a ThermoFisher ICAP 6000 Duo inductively coupled plasma optical emission spectrometer. Three measurements were performed on each sample to obtain an averaged value, and three different synthesis batches were measured to yield an average Ag:S stoichiometry of (1.4 ± 0.1) : (1). Being substoichiometric, the Ag$_{2-x}$S crystal favours memory effects by increasing the electron conductivity.[3]

*X-Ray diffraction* (XRD) analysis was performed with a PANanalytical Empyrean X-ray diffractometer equipped with a 1.8 kW CuKα ceramic X-ray tube, PIXcel[3D] 2x2 area detector and operating at 45 kV and 40 mA. Samples for the XRD measurements were prepared in a glove box by drop casting a concentrated Ag$_{2-x}$S NC dispersion onto a miscut silicon wafer. The diffraction patterns were collected using a Parallel-Beam (PB) geometry and symmetric reflection mode. According to the XRD data, the crystal structure of the colloidal Ag$_{2-x}$S NCs corresponds to the acanthite (α-Ag$_2$S) phase together with some cubic metallic Ag inclusions (**Fig. S2a**). Note that, although the β-Ag$_2$S (argentite) phase presents superionic properties, it is only stable at temperatures higher than 177ºC.[4,5] We have tried to obtain β-Ag$_2$S, however, neither synthesis at 250ºC (**Fig. S2b**) nor annealing an α-Ag$_2$S film at 250ºC (**Fig. S2c**) yielded the β-phase crystal structure (**Fig. S2d**). Likely, during the cooling process a relaxation of the crystal structure from argentite to acanthite occurs.

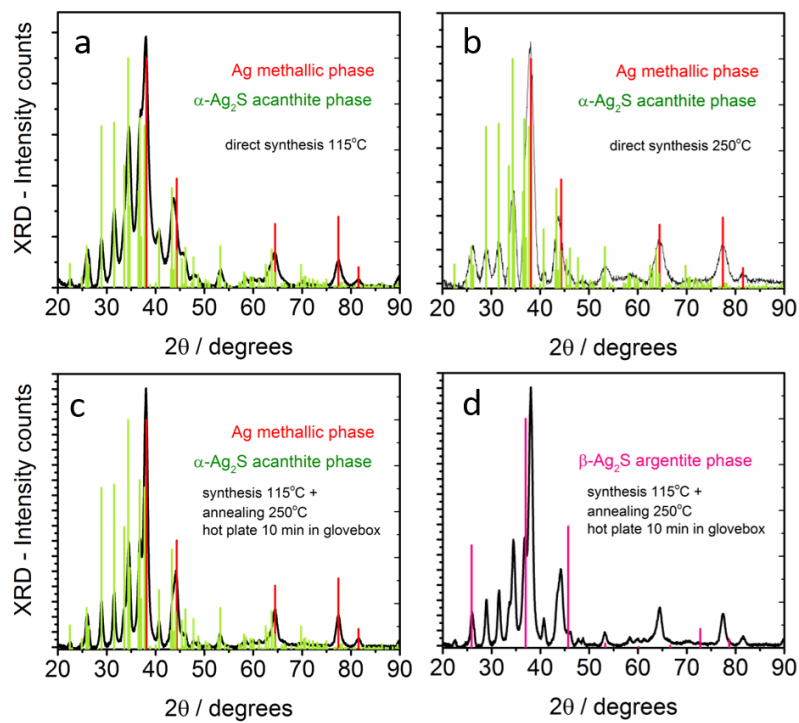

**Fig. S2.** XRD spectra of Ag$_{2-x}$S NCs samples from direct synthesis at 115°C (a) and 250°C (b) together with the results from annealing at 250°C in inert atmosphere of sample (a) (c, d). The XRD reference patterns of metallic silver (cubic, code 98-042-6921), acanthite Ag$_2$S (monoclinic, code 98-003-0445) and argentite Ag$_2$S (cubic, code 98-000-9586) are from the ICSD database and shown as vertical lines.

**Additional optical and SEM images for Ag/Ag$_{2-x}$S/Ag macroscale system.**

To complement the information of Fig. 1 in the main text, in **Fig. S3** we included detailed images collected with optical microscope and SEM, in which it is possible to observe the formation of the Ag metallic protrusions at the Ag negative electrode (noted as cathode for ECM devices[6]) and an Ag depletion at the anode (**Fig. S3b**). **Fig. S3c** also highlights how the Ag protrusions growth follows the direction of the electrical field.

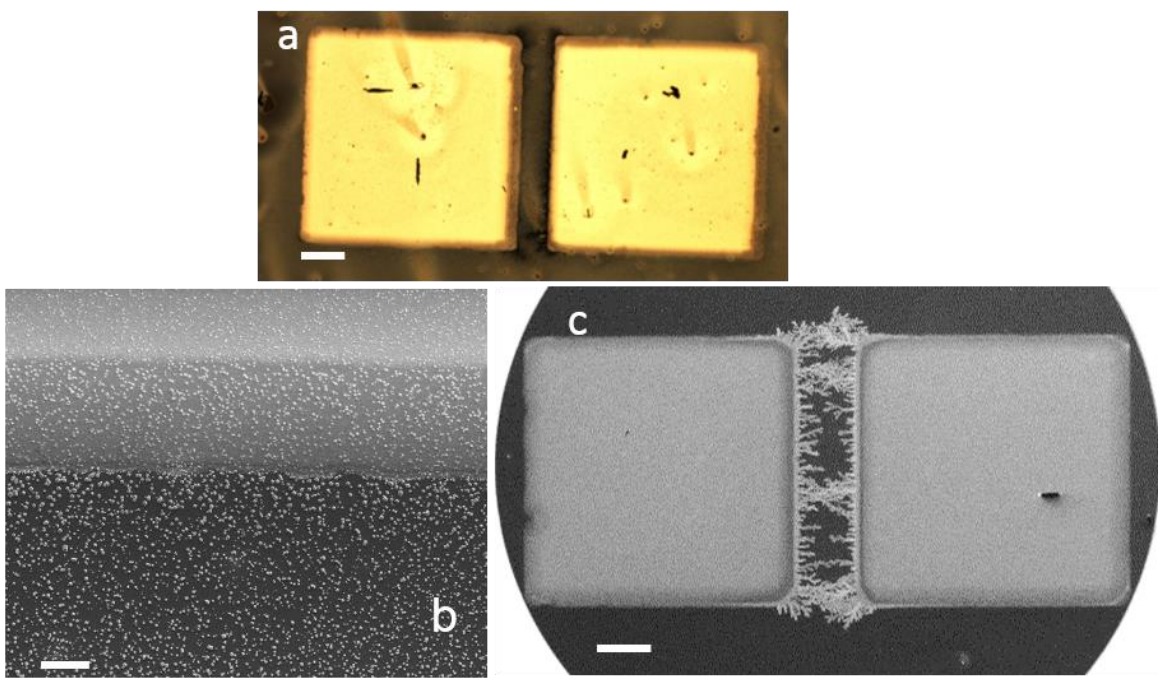

**Fig. S3.** (a) Optical image of a representative Ag/Ag$_{2-x}$S/Ag device. Scale bar is 200 µm. (b, c) SEM images of a representative Ag/Ag$_{2-x}$S/Ag device with film thickness of 20 nm, showing the area of the Ag contact where it is possible to observe the Ag depletion and the state of the device after several DC voltage sweeps. As shown in the SEM image in (c), dendrites are formed along the electric field lines. Scale bars are 10 (b) and 200 (c) µm, respectively.

**SEM/EDX analysis of Ag/Ag$_{2-x}$S/Ag macroscale system.**

Complementary SEM/EDX analysis is shown in **Fig. S4** for a representative Ag/Ag$_{2-x}$S/Ag macroscale device. The EDX mapping shows the accumulation of Ag in the protrusions, while the S signal remains constant along the film.

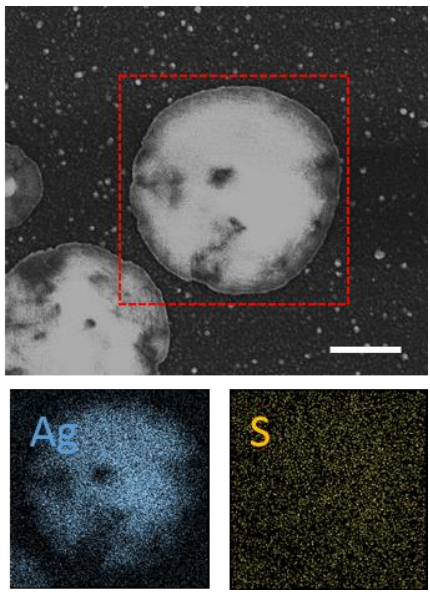

**Fig. S4.** SEM image of a single Ag protrusion. Red box marks the area analysed by EDX mapping. Scale bar is 500 nm. The lower panels show EDX maps acquired in the area marked in main panel for Ag (Lα1, 2.98 keV) and S (Kα1, 2.31 keV) lines, respectively.

**Control experiments: vacuum and high voltage**

Control experiments under vacuum demonstrated that protrusions with similar morphology to the ones in air are formed, **Fig. S5a-b,** indicating that the reaction is happening with the sulphur rather than the oxygen (or moisture) in air. Moreover, the IV characteristics collected under vacuum show the same features (within the variability we observed between different samples and devices) in comparison with the ones in air, indicating that no more species are involved in the switching mechanism, **Fig. S5c-d**. Additionally, the endurance is not improved in vacuum (**Fig. S5e**).

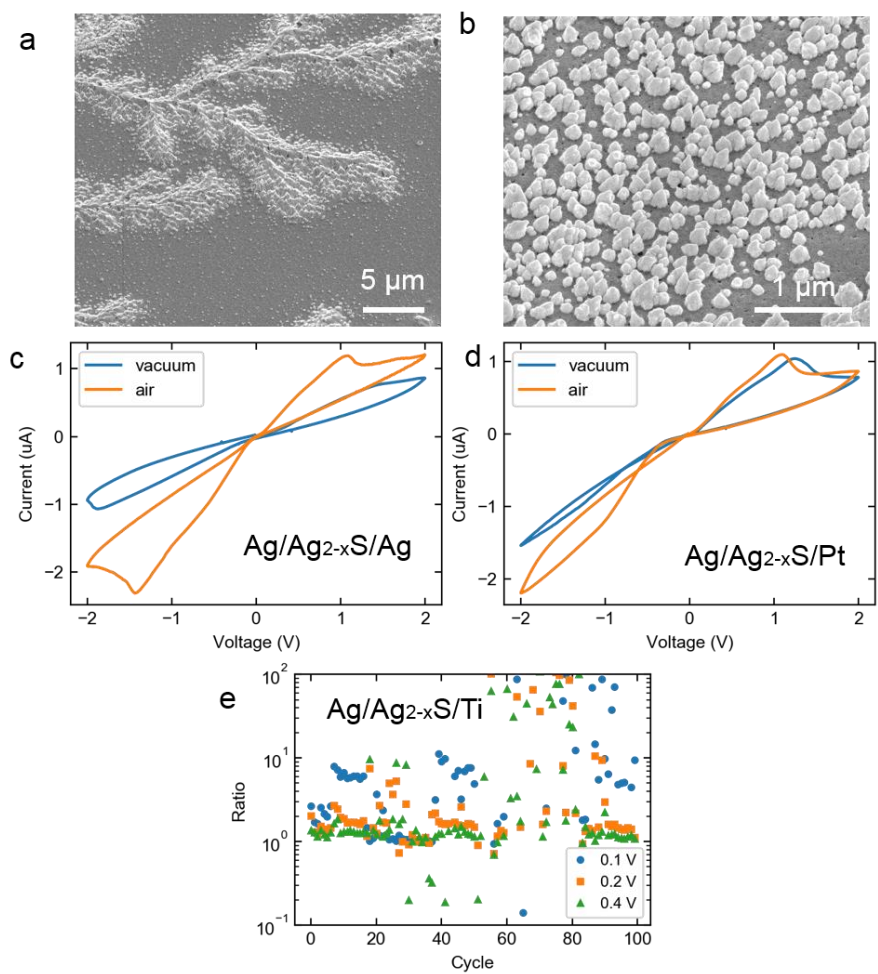

**Fig. S5.** Representative SEM images for Ag/Ag$_{2-x}$S/Ag (a) and Ag/Ag$_{2-x}$S/Pt (b) 'macroscale' devices after electrical measurements performed under vacuum. IV characteristics collected in vacuum and in air conditions for Ag/Ag$_{2-x}$S/Ag (c) and d) Ag/Ag$_{2-x}$S/Pt (d) 'macroscale' devices. The measurements in air are the same reported in the main text. (e) Multi-read' experiment in an Ag/Ag$_{2-x}$S/Ti 'macroscale' device in vacuum. All the experiments were performed as described in the main text of the paper, except for the use of a probe station under vacuum (pressure ~10$^{-5}$ mbar) for the measurements.

We also measured cycles starting at 0V, up to 5V, down to -5 V and back to 0 V for representative Ti/Ag$_{2-x}$S/Ag analogous to Fig. 3a (main text) and Ag/Ag$_{2-x}$S/Ag devices, under air and vacuum conditions (**Fig. S6).** The trend for the first loop (positive) is very similar to cycles up to 2 V, but when the voltage becomes negative (approximately from -2V), the formation of a large metallic (Ag) dendritic structure contacts the two electrodes and results in a very high current (denoted by a sudden vertical line). The formation of the large dendritic structure is mostly irreversible since the filament remains and keeps short-circuiting the device.

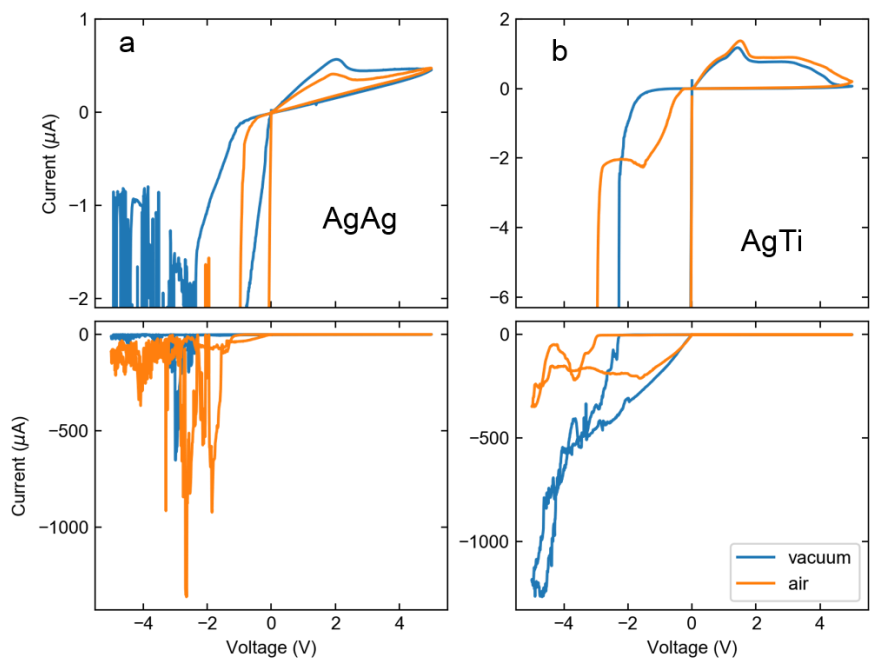

**Fig. S6.** Representative IV cycles up to 5V in 'macroscale' devices, collected in vacuum (blue) and in air (orange) conditions, for Ag/Ag$_{2-x}$S/Ag (a) and Ag/Ag$_{2-x}$S/Ti (b) systems. The lower panels show the upper curves at full scale.

**Pre-activation protocols and switching tests.**

As mentioned in the main text, to stabilize the first set/read/reset cycles reading, we established a pre-activation protocol. In **Fig. S7a**, we show a scheme of the procedure for the initial formation operation (0 → 2V) and the subsequent 'multi-read' protocol. **Fig. S7b-d** include pre-activated and non-pre-activated data in different devices, highlighting how the initiation of the ion migration across the $Ag_{2-x}S$ film by the initial voltage ramp reduces the number of cycles required to reach a stable measurement.

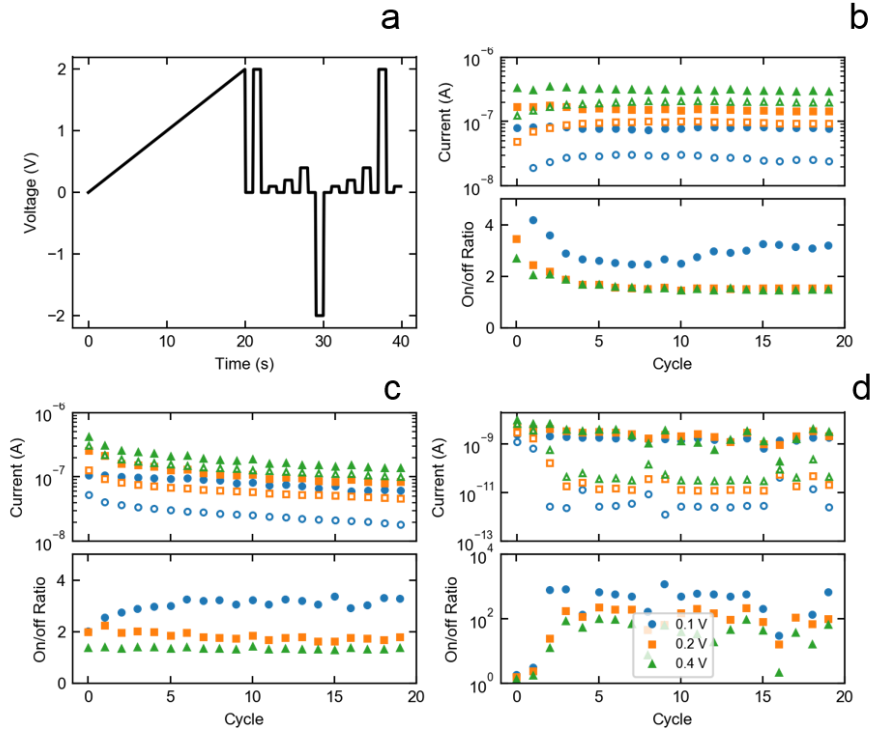

**Fig. S7.** (a) Pre-activation procedure (voltage as a function of time): after a linear ramp from 0 to 2 V, we start the set/read/reset cycles. (b-d) Effect of the activation on set/reset cycles. Upper panels: ON (full symbols) and OFF (empty symbols) current for set/reset cycles measured with $V_{read}$= 0.1 V (●), 0.2 V (■) and 0.4 V (▲). The $I_{ON}/I_{OFF}$ ratio at the different read voltages is reported in the lower panels. (b) $Ag/Ag_{2-x}S/Ag$ macro-scale device, pre-activated with the procedure in panel (a). The ratio stabilizes to 3.0±0.1 after 4 cycles. (c) $Ag/Ag_{2-x}S/Ti$ macro-scale device, not pre-activated, reaches a stable ratio after 4 cycles (for a pre-activated device, see Fig. 3c in the main text). (d) Ti/Ag, tip/tip micro-scale device fabricated by EBL. The device was not pre-activated before the set/reset cycles, and shows a significant increase in $I_{ON}/I_{OFF}$ ratio during the first 4 read cycles.

**SEM/EDX analysis of Ti/Ag$_{2-x}$S/Ag macroscale system.**

Complementary SEM/EDX analysis is shown in **Fig. S8** for a representative macroscale device with inert electrode: Ag/Ag$_{2-x}$S/Ti. The EDX maps demonstrate that no diffusion of Ti occurs during the device operation.

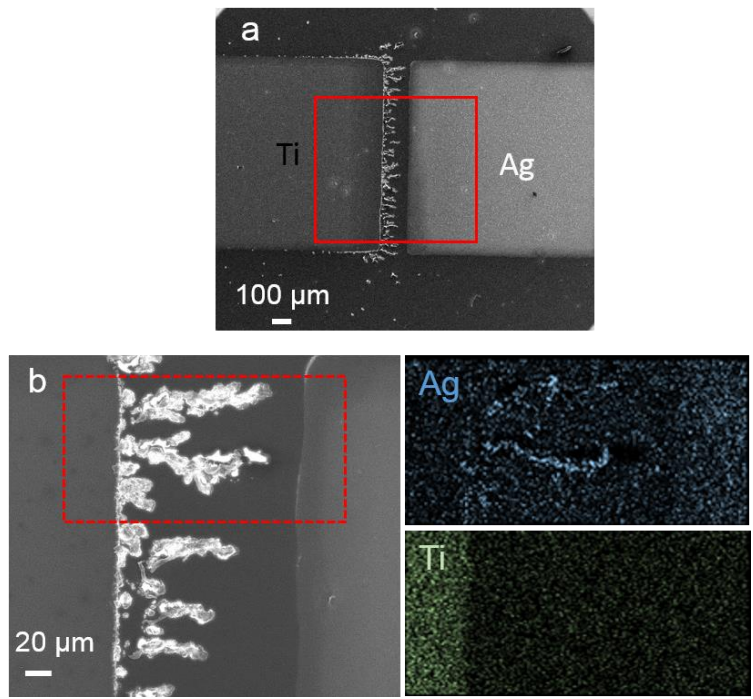

**Fig. S8.** (a, b) SEM images of a representative Ti/Ag$_{2-x}$S/Ag device with two different magnifications, showing the dendrite formation at the Ti electrode. The red box in (a) marks the area shown in **Fig. 3a** in the main text. EDX mapping analysis of the dendrites is included in (b). The dashed box marks the area analysed by EDX; the panels on the right show the intensity maps for Ag (Lα1, 2.98 keV) and Ti (Kα1, 4.51 keV) lines.

**Switching tests on different electrode geometries prepared by EBL**

Apart from the data shown in the main text, **Fig. 4**, focused on the tip-tip electrode geometry, here we include results for the other geometries explored (**Fig. S9**). A representative example for $I_{ON}/I_{OFF}$ ratio, obtained by applying the 'multi-read protocol', for a flat-flat configuration in a Pt/Ag$_{2-x}$S/Ag device is shown in **Fig. S9a**, resulting in a best ratio of 115±28 ($V_{read}$ = 0.2-0.4V), while results for a representative Ti/Ag$_{2-x}$S/Ag tip-flat device are shown in **Fig. S9b**, demonstrating a best ratio of 651±139 ($V_{read}$ = 0.1V).

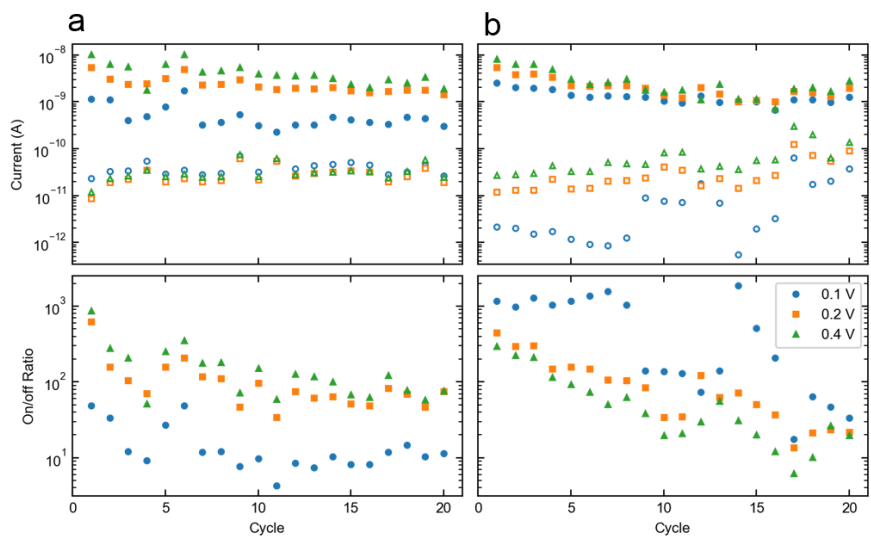

**Fig. S9.** Set/reset cycles measured with $V_{read}$ = 0.1 V (●), 0.2 V (■) and 0.4 V (▲) for (a) flat(Pt)/flat(Ag) and (b) tip(Ti)/flat(Ag) devices with a 20 µm gap. $I_{ON}$ (full symbols) and $I_{OFF}$ (empty symbols) for both are displayed in the upper panels, and the $I_{ON}/I_{OFF}$ ratio in lower panels. Ag was used as cathode (-); devices were pre-activated before measurements.

Focusing on the best tip-tip configuration, Ag/Ag$_{2-x}$S/Ti device, we performed set/read/reset test up to 100 cycles in different samples, **Fig. S10**, verifying that we obtained reproducible switching of the EBL fabricated devices only up to 20 cycles.

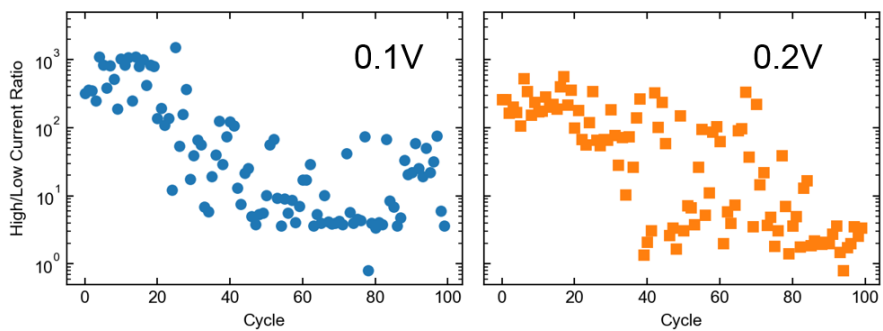

**Fig. S10**. Evolution of the $I_{ON}/I_{OFF}$ ratio at $V_{read}$ = 0.1 (right) and 0.2 V (left) for two different tip-tip Ag/Ag$_{2-x}$S/Ti devices stressed during 100 cycles of SET/RESET (±2 V) and read in air.

**Supplementary Movies**

The movies were acquired with a stereomicroscope (Olympus® SZX7) coupled to a CCD camera (Olympus® UC30). The original resolution of the video was 1040x772 pixels, acquired without compression; the files present in the supplementary were compressed, and finally overlays were added to assist the interpretation.

*Movie 1*

Movie 1 was acquired on an Ag/Ag$_{2-x}$S/Ag macroscale device (pad 1x1 mm$^2$) during a voltage sweep from 0 V to +2 V, to -2 V and finally back to 0 V analogous to **Fig. 2a** in the main text. The bottom electrode is connected to ground (cathode (-)), while the voltage was applied to the top electrode (anode (+)). The voltage is shown in the video only for selected values during the linear sweep of the measurements. Red arrows mark the formation of dendrites and the depletion of Ag in the electrode.

*Movie 2*

Movie2 2 was acquired during pulse cycles with high voltage value of +5 V and -5 V of 1 s duration, to speed up the Ag dendrite growth, in an Ag/Ag$_{2-x}$S/Ag macroscale device (pad 1x1 mm$^2$). The bottom electrode is connected to ground (cathode (-)), while the voltage was applied to the top electrode (anode (+)). In the video the labels "+" and "-" symbols identify where the positive and negative potential is applied at each step. It is possible to observe that the dendrites form from the negative electrode.

**Supplementary references**

(1) Doh, H.; Hwang, S.; Kim, S. Size-Tunable Synthesis of Nearly Monodisperse Ag$_2$S Nanoparticles and Size-Dependent Fate of the Crystal Structures upon Cation Exchange to AgInS$_2$ Nanoparticles. *Chem. Mater.* **2016**, *28* (22), 8123–8127.
(2) Lin, S.; Feng, Y.; Wen, X.; Zhang, P.; Woo, S.; Shrestha, S.; Conibeer, G.; Huang, S. Theoretical and Experimental Investigation of the Electronic Structure and Quantum Confinement of Wet-Chemistry Synthesized Ag$_2$S Nanocrystals. *J. Phys. Chem. C* **2015**, *119* (1), 867–872.
(3) Riess, I. Mixed Ionic–Electronic Conductors—Material Properties and Applications. *Solid State Ion.* **2003**, *157* (1–4), 1–17.
(4) Lutz, C.; Hasegawa, T.; Chikyow, T. Ag$_2$S Atomic Switch-Based 'Tug of War' for Decision Making. *Nanoscale* **2016**, *8* (29), 14031–14036.

(5)  Sadovnikov, S. I.; Gusev, A. I.; Rempel, A. A. An in Situ High-Temperature Scanning Electron Microscopy Study of Acanthite–Argentite Phase Transformation in Nanocrystalline Silver Sulfide Powder. *Phys. Chem. Chem. Phys.* **2015**, *17* (32), 20495–20501.

(6)  Kozicki, M. N.; Mitkova, M.; Valov, I. Electrochemical Metallization Memories. In *Resistive Switching*; Ielmini, D., Waser, R., Eds.; Wiley-VCH Verlag GmbH & Co. KGaA: Weinheim, Germany, 2016; pp 483–514.12